\documentclass[twocolumn,amsmath,amssymb]{revtex4}
\usepackage{graphicx}% Include figure files
\usepackage{dcolumn}% Align table columns on decimal point
\usepackage{bm}% bold math
\usepackage{amsthm}

\def\ket#1{| #1 \rangle}
\def\bra#1{\langle #1 |}

\def\norm#1{\| #1 \|}

\def\diag{\operatorname{diag}}

\def\Span{\operatorname{span}}

\def\Tr{\operatorname{Tr}}

\def\C{\mathcal{C}}

\def\S{\mathcal{S}}

\begin{document}
\title{Entanglement generation between distant atoms by Lyapunov control}
\author{Xiaoting Wang}\email{xw233@cam.ac.uk}
\affiliation{Department of Applied Maths and Theoretical Physics,
             University of Cambridge, Wilberforce Road, Cambridge, CB3 0WA, UK}
\author{S.~G.~Schirmer}\email{sgs29@cam.ac.uk}
\affiliation{Department of Applied Maths and Theoretical Physics,
             University of Cambridge, Wilberforce Road, Cambridge, CB3 0WA, UK}
\date{\today}

\begin{abstract}
We show how to apply Lyapunov control design to the problem of
entanglement creation between two atoms in distant cavities
connected by optical fibers.  The Lyapunov control design is optimal
in the sense that the distance from the target state decreases
monotonically and exponentially, and the concurrence increases
accordingly. This method is far more robust than simple geometric
schemes.
\end{abstract}
\maketitle

\section{Introduction}
\label{sec:intro}

Atoms, or their artificial counterparts, quantum dots, in cavities
or traps have great potential for applications in quantum
communication, metrology and information processing.  Since
entanglement is a crucial resource in quantum computation and
communication, the preparation of maximally entangled states is a
crucial task.  Nonlocal interactions between two physical qubits are
required to generate entanglement and there have been numerous
proposals to effect such interactions, especially for atoms trapped
in distant cavities~\cite{Cirac,Pellizzari,
van,Sorensen,Bose,Parkins,Mancini,Duan1,Duan2,Simon,Clark,Browne,
Mancini-Bose,Wang-Mancini}, and similar schemes are conceivable for
artificial atoms such as quantum dots.  Some of these proposals make
use of continuous feedback in open quantum
systems~\cite{Wang-Mancini} but most are based on Hamiltonian
systems, and in most cases only simple geometric control schemes are
employed to create the maximally entangled state. These methods have
the advantage of simplicity but unfortunately often suffer from
robustness issues.

In this work we explore an alternative control design inspired by
Lyapunov
functions~\cite{Vettori,Ferrante,Grivopoulos,Mirrahimi2004a,
Mirrahimi2004b,Mirrahimi2005,altafini1,altafini2,Wang-Schirmer,ENOC08}
to find control designs for robust entanglement creation.  Lyapunov
control design has the advantage of being sufficient simple to be
amenable to rigorous analysis, and much is known about their
convergence properties, robustness and stability.  In particular
such design can be shown to be highly effective for systems that
satisfy certain sufficient conditions, which are roughly equivalent
to the controllability of the linearized
system~\cite{Mirrahimi2005,altafini2}. Unfortunately, this appears
to be a strong requirement not satisfied by many physical systems.
However, in certain cases, in particular for systems like the
two-atom model proposed by Mancini and Bose~\cite{Mancini-Bose}, we
can circumvent these restrictions by considering the dynamics on a
subspace and successfully apply Lyapunov control to create maximally
entangled states from certain initial product states in robust
fashion.

The paper is organized as follows.  In Sec.~\ref{sec:model} we
briefly review the distant-atom model and the geometric control
scheme proposed in \cite{Mancini-Bose} to generate entanglement.  In
Sec.~\ref{sec:Lyapunov} we briefly review Lyapunov control and show
how to apply it to the problem of steering the system from certain
product state to one of the four Bell state in a robust fashion. We
will consider two control paradigms: one is to control the local
Hamiltonian which is easier to implement experimentally; the other
is to control the non-local interaction Hamiltonian, which might be
possible for certain systems.

%Finally, in Sec. IV, we analyze the robustness of the scheme with
%regard to various errors and compare the current proposal with the
%original geometric control design.

\section{Two-distant-atom model and geometric control}
\label{sec:model}

\begin{figure}
\includegraphics[width=\columnwidth]{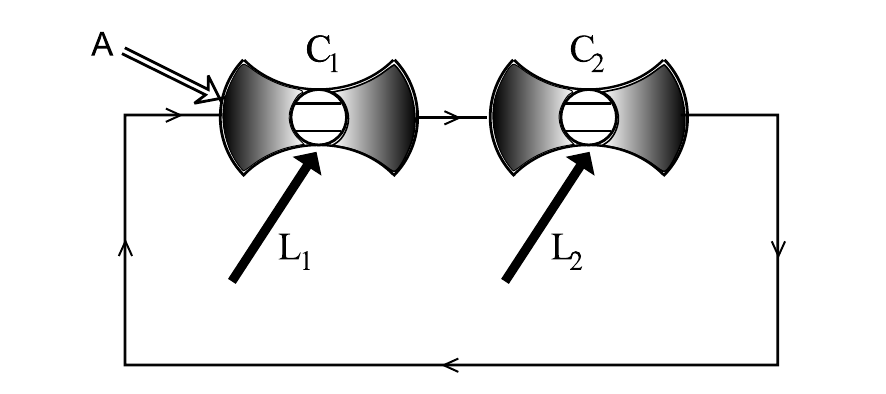}
\caption{Two cavities $C_1$ and $C_2$, each of which contains a
two-level atom, are connected into a closed loop through optical
fibers. The off-resonant driving field $A$ generates an effective
non-local Hamiltonian $H_{eff}$ while the two local resonant lasers
generate the local Hamiltonian $H_{local}$.} \label{fig0}
\end{figure}

We consider a two-qubit model where the qubits are encoded in two
atoms or two quantum dots in distant cavities connected into a
closed loop by optical fibers, as illustrated in Fig.~\ref{fig0}. It
was shown in~\cite{Mancini-Bose} that eliminating the interacting
light field between the two atoms in the dispersive regime leads to
an effective Hamiltonian for the two-atom system of form $H_{\rm
tot}=H_{\rm local}+H_{\rm eff}$, where the local Hamiltonian induced
by interaction with resonant light and the effective interaction
Hamiltonian are:
\begin{subequations}
 \label{eqn:two-atom}
\begin{align}
  H_{\rm local} &= B (X\otimes I+ I\otimes X)\\
  H_{\rm eff} &= 2 J Z\otimes Z
\end{align}
\end{subequations}
where $X$, $Y$, $Z$ are Pauli operators and $I$ is the identity
operator, and the coupling constant $B=\eta J$ where $\eta$ should
be sufficiently smaller than $1$ to ensure the derivation of $H_{\rm
eff}$ remains valid.

\begin{figure}
\includegraphics[width=\columnwidth]{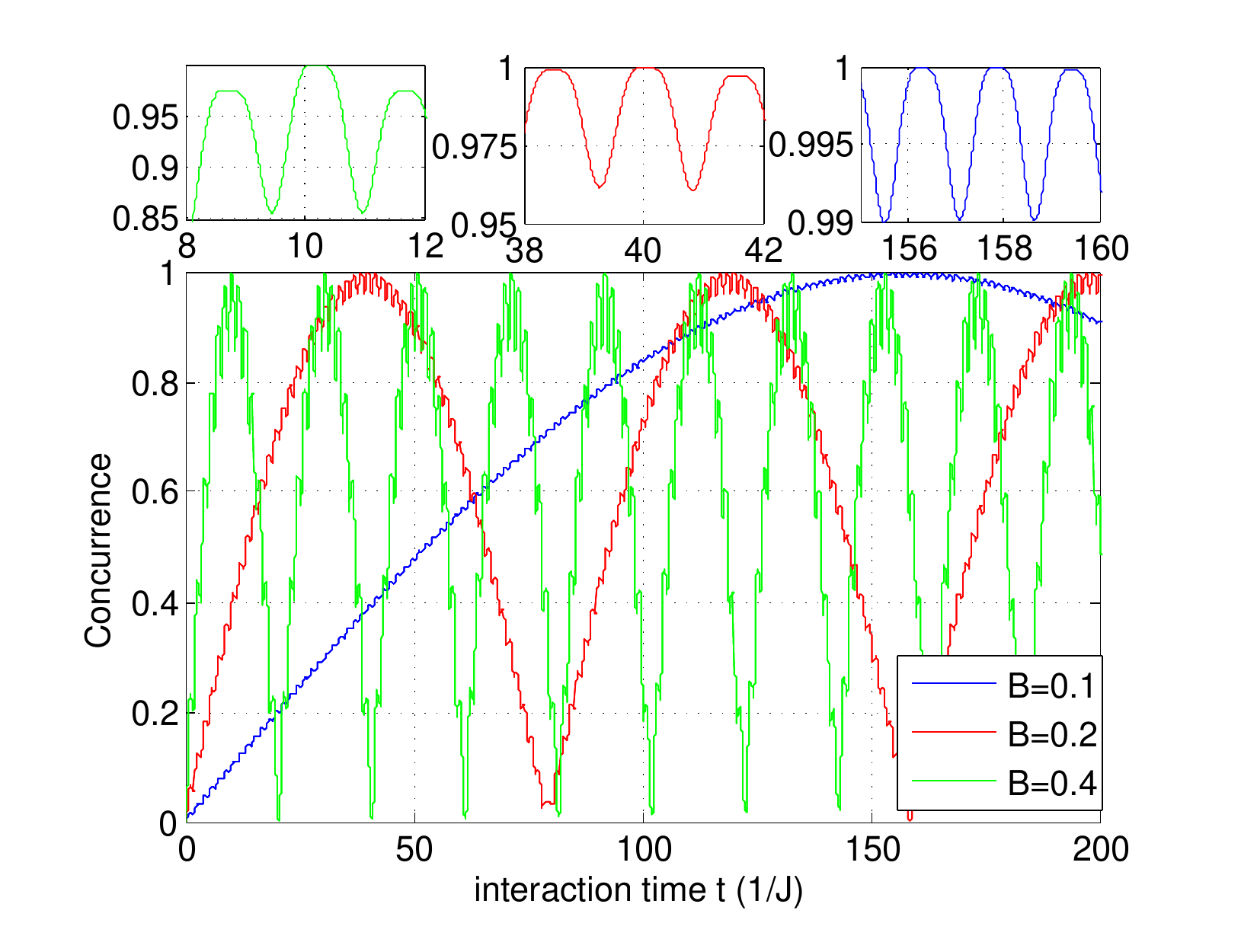}
\caption{Concurrence as a function of the interaction time for the
geometric control scheme for different values of coupling $B$.
Achieving unit concurrence requires requires switching off the
Hamiltonian precisely, due to the fluctuations concurrence curve. If
the control Hamiltonian is switched on or off too early or too late,
even by a small amount, the concurrence of the final state may be
reduced significantly. The three sub-figures on top of the main
figure shows the zoom-in of the plots.} \label{fig1}
\end{figure}

This Hamiltonian can be used to generate a maximally entangled state
from the initial ground state by turning on $H_{\rm tot}$ for a
critical time $t_0$ before switching the field
off~\cite{Mancini-Bose}.  Broadly speaking, by applying a constant
Hamiltonian we effectively perform a rotation about a fixed axis in
the two-qubit space, and with the correct timing we can choose the
rotation angle such as to ensure that the system state ends at the
correct target state.  However, plotting the concurrence of the
final state versus the interaction time (Fig.~\ref{fig1}) shows that
achieving very high fidelity with respect to the maximally entangled
state requires very precise switching as the concurrence is subject
to small fluctuations.  In the model we have assumed a fixed
coupling strength $J$ and controllable local field $B$. We see that
increasing $B$ significantly reduces the time required to prepare a
maximally entangled state but also increased the magnitude of the
fluctuations.  E.g., for $B=0.1$ the fluctuations around the peak
are only about $1$\% but it takes $157$ time units to reach a
maximally entangled state.  For $B=0.4$ on the other hand, we can
prepare a maximally entangled state in about $1/8$ of the time but
the concurrence fluctuations increase by a factor of approximately
$15$. Therefore, although such design for entanglement generation is
quite simple, it is not robust against imperfections of switching
time.

\section{Lyapunov Control Design}
\label{sec:Lyapunov}

In the previous section we have seen that the method of entanglement
generation by switching a constant field on for a fixed amount of
time is highly sensitive to small switching time errors. Ideally, we
would like a control scheme where the concurrence of the two qubits
converges to $1$ asymptotically, and better without any
fluctuations. In that way, the control is robust against switching
time errors. A simple method that seems well suited to this task is
Lyapunov-based design. Roughly speaking, the idea of Lyapunov
control is to choose a suitable so-called Lyapunov function $V$ and
then try to find a control that ensures that $V$ is monotonically
decreasing along any dynamical evolution.

In the time scale where the Hamiltonian evolution is still a good
approximation, many physical systems satisfy the the quantum
Liouville equation (with $\hbar=1$)
\begin{align*}
\dot \rho = -i[H_0+f(t)H_1,\rho],
\end{align*}
where we have assumed the Hamiltonian has two parts: $H_0$ is the
system Hamiltonian and $H_1$ is the interaction Hamiltonian, with
the interaction coupling constant modulated by the function $f(t)$.
For example, for a two-level structure of a single atom,
$H_0=\frac{\Omega}{2}\sigma_z$ is the energy splitting, and
$H_1=\sigma_x$ is the dipole interaction between the laser and the
atom, with a varying $f(t)$ by modulating the laser amplitude. The
fact that $f(t)$ can be varied is very crucial from control point of
view, since this degree of freedom allows us to design the dynamics
to derive the desired evolution.

We can define a general control task thus: for a given target state
$\rho_d$, for example, a maximally entangled state, we wish to find
a control function $f(t)$, such that the system state $\rho(t)$ will
converge to $\rho_d$, as $t\to\infty$. In many applications, we
allow $\rho_d(t)$ to evolve under $H_0$, and the control requirement
becomes $\rho(t)\to\rho_d(t)$ as $t\to\infty$, which is generally
known as tracking control~\cite{Bohacek}. In the following we
assume:
\begin{equation*}
 \dot \rho_d = -i[H_0,\rho_d]
\end{equation*}

Motivated from the theory of Lyapunov function and the Hilbert
Schmidt distance $\norm{\rho(t)-\rho_d(t)}_2$, we define
\begin{equation}
 V(\rho,\rho_d) = \frac{1}{2}\norm{\rho-\rho_d}^2
                = \frac{1}{2}\Tr[(\rho-\rho_d)^2].
\end{equation}

Assuming $\kappa>0$, if we choose
\begin{equation}\label{eqn:f}
  f(t) = f(\rho(t),\rho_d(t)) = \kappa \Tr(\rho_d(t)[-iH_1,\rho(t)]),
\end{equation}
we find that for $V(t)=V(\rho(t),\rho_d(t))$,
\begin{equation}
  \dot V(t)=-f(t)\Tr(\rho_d(t)[-iH_1,\rho(t)])=-\kappa f(t)^2\le 0.
\end{equation}
Hence $V$ is a Lypunov function and the value of $V$ monotonically
decreases along any solution $(\rho(t),\rho_d(t))$. Moreover, every
solution $(\rho(t),\rho_d(t))$ converges to an invariant set $E$,
called the LaSalle invariant set, on which $\dot{V}$ vanishes.

Discussions on Lyapunov-based design in terms of density operators
have been analyzed~\cite{altafini1,altafini2, Wang-Schirmer}. In
particular, many target states can be shown to be almost globally
asymptotically stable if the Hamiltonian satisfies certain demanding
conditions: (i) $H_0$ be strongly regular and (ii) $H_1$ be fully
connected~\cite{Wang-Schirmer}. The former condition translates into
the requirement that $H_0$ have distinct transition frequencies
between any pair of energy levels. This rules out systems with
degenerate or equally spaced energy levels. The latter condition is
even more demanding. In the basis where $H_0$ is diagonal, all the
off-diagonal elements of $H_1$ must be non-zero, i.e. transitions
between any two energy level of $H_0$ can be realized. When the
strict conditions on the Hamiltonian do not hold, for most cases,
the target state can be shown to be no longer asymptotically stable,
and we no longer have $\rho(t)\to\rho_d(t)$, implying that the
control design becomes ineffective. This really restricts the
applicability of the method especially for higher-dimensional
systems, including two-qubit models and spin chains.

However, for high-dimensional systems with Hamiltonian not
satisfying the above conditions, it is still possible to make the
target state asymptotically stable on a subspace, where the Lyapunov
control can be applied effectively. In the following, for the
two-distant-atom model (Fig.~\ref{fig0}), we illustrate how the
Lyapunov control design can be utilized to drive the system state
from a product state to a maximally entangled state, despite the
fact that the full Hamiltonian of the system clearly does not
satisfy the strict conditions set out above.

\section{Lyapunov Control Design for Entanglement Creation}
\label{sec:entanglement}

For the two-distant-atom model with Hamiltonian
(\ref{eqn:two-atom}), we can either choose the control Hamiltonian
$H_1$ to be the local Hamiltonians $H_1=H_{\rm local}$ or the
effective coupling Hamiltonian $H_1=H_{\rm eff}$, depending on which
scenario is easier to implement for a particular physical system.

\subsection{Local Control}

\begin{figure}
\includegraphics[width=\columnwidth]{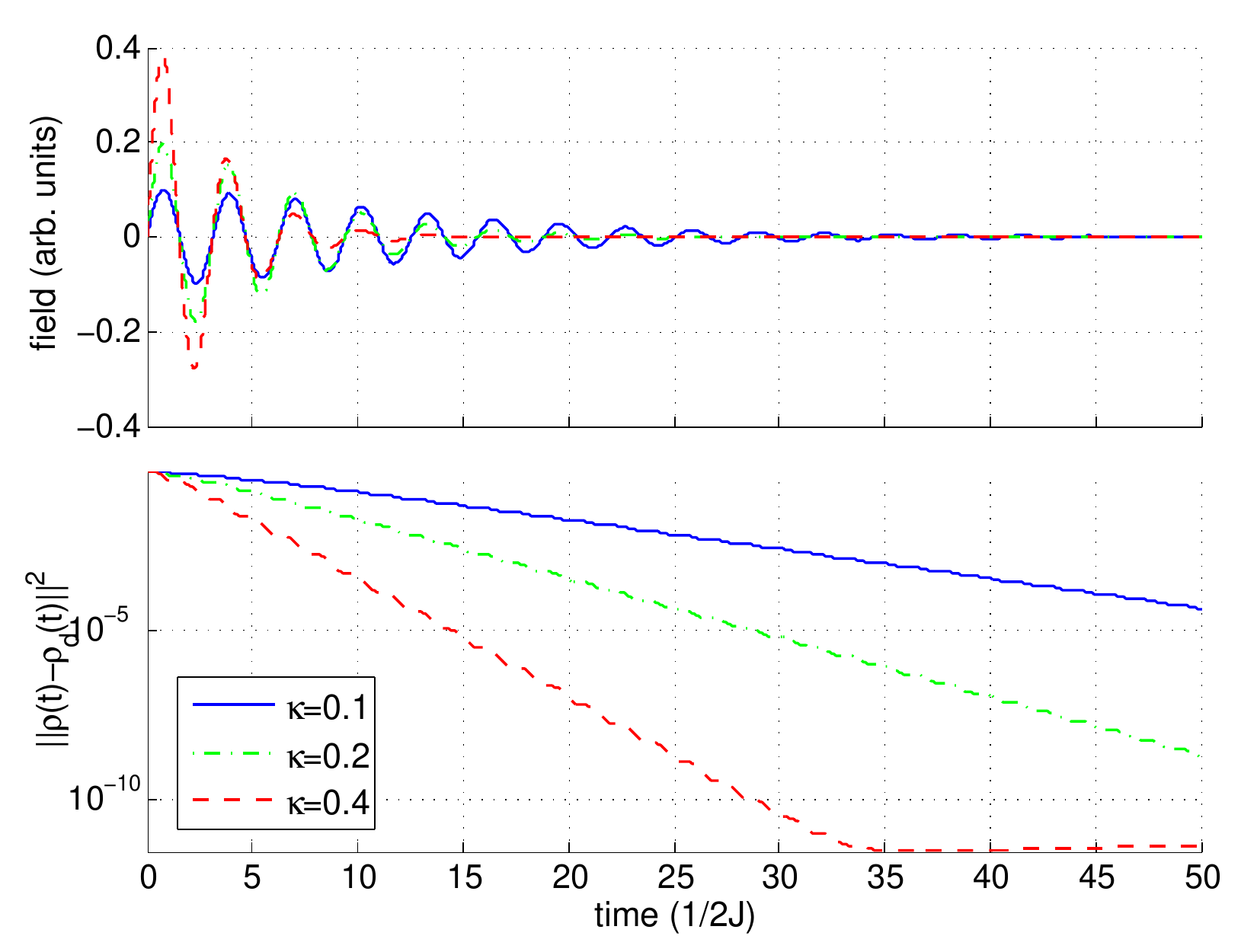}
\caption{Local Control: control fields obtained from Lyapunov design
for different values of $\kappa$ and distance between the system
state and the Bell state $\ket{\Psi^+}$.  The control design is
robust in that the field amplitude gently decreases to zero, and the
semilog distance plot shows that the convergence to the target state
is not only monotonic but also exponential with the converging rate
determined by $\kappa$.} \label{fig2}
\end{figure}

First, let us consider controlling the local Hamiltonian.  In this
case we choose $H_0=H_{\rm eff}=2J(Z\otimes Z)$ and $H_1=H_{\rm
local}= \eta J(X\otimes I+ I\otimes X)$.  To make the Hamiltonian
easier to analyze, we transform from the $Z$-eigenbasis
$\{\ket{0},\ket{1}\}$ to the $X$-eigenbasis $\{\ket{+},\ket{-}\}$.
In this basis, the matrices for the Hamiltonian are rewritten as

\begin{align*}
H_0=2J\begin{pmatrix}
  0 & 0 & 0 & 1\\
  0 & 0 & 1 & 0\\
  0 & 1 & 0 & 0\\
  1 & 0 & 0 & 0
\end{pmatrix}, \quad
H_1=2\eta J
\begin{pmatrix}
    1 & 0 & 0 & 0\\
    0 & 0 & 0 & 0\\
    0 & 0 & 0 & 0\\
    0 & 0 & 0 & -1
\end{pmatrix}.
\end{align*}
and it is easy to see that the eigenvectors of $H_0$ are the Bell
states
\begin{subequations}
\begin{align}
\ket{\Psi^+} &= \frac{1}{\sqrt{2}}(\ket{+-}+\ket{-+})\\
\ket{\Phi^+} &= \frac{1}{\sqrt{2}}(\ket{++}+\ket{--})\\
\ket{\Phi^-} &= \frac{1}{\sqrt{2}}(\ket{++}-\ket{--})\\
\ket{\Psi^-} &=\frac{1}{\sqrt{2}}(\ket{+-}-\ket{-+}).
\end{align}
\end{subequations}

To generate maximally entangled state, we can choose
$\rho_d=\ket{\Phi^+}\bra{\Phi^+}$, for instance, and the control
$f(t)=\kappa\Tr(\rho_d[-iH_1,\rho(t)])$, according to (\ref{eqn:f}).
Notice that $H_0$ and $H_1$ do not satisfy the strict condition in
Section~\ref{sec:Lyapunov}. Thus this design cannot drive every
state to the target state, but we can see that if the initial state
of the system is $\rho(0)=\ket{++}\bra{++}$ or $\ket{--}\bra{--}$
then the state will converge to the target state. In fact, in the
Bell-state basis, the Hamiltonian can be written as
\begin{align*}
\tilde{H}_0=2J\begin{pmatrix}
  1 & 0 &  0 &  0\\
  0 & 1 &  0 &  0\\
  0 & 0 & -1 &  0\\
  0 & 0 &  0 & -1
\end{pmatrix}, \;
\tilde{H}_1= 2\eta J\begin{pmatrix}
0 & 0 & 0 & 0 \\
0 & 0 & 1 & 0\\
0 & 1 & 0 & 0\\
0 & 0 & 0 & 0
\end{pmatrix},
\end{align*}
where
\begin{align*}
\rho_d=\begin{pmatrix}
0 & 0 & 0 & 0 \\
0 & 1 & 0 & 0\\
0 & 0 & 0 & 0\\
0 & 0 & 0 & 0
\end{pmatrix}, \quad
\rho(0)=\frac{1}{2}\begin{pmatrix}
0 & 0 & 0 &  0 \\
0 & 1 & \pm1 &  0\\
0 & \pm1 & 1 &  0\\
0 & 0 & 0 &  0
\end{pmatrix}.
\end{align*}

For states initially prepared in the subspace $\S$ spanned by
$\ket{++}$ and $\ket{--}$, we clearly see that the dynamics under
the Hamiltonian $H_0+f(t)H_1$ will be confined in that subspace, and
thus we can consider the dynamics on this two-dimensional subspace
$\S$ where the Hamiltonians and state take the form:
\begin{align*}
H_0    = 2J \begin{pmatrix} 1 & 0 \\ 0 & -1 \end{pmatrix}, \; H_1 =
2\eta J \begin{pmatrix} 0 & 1 \\ 1 &  0 \end{pmatrix}, \; \rho_d =
\begin{pmatrix} 1 & 0 \\ 0 &  0 \end{pmatrix}.
\end{align*}
The results in~\cite{Wang-Schirmer} now guarantee that all solutions
in $\S$ except for $\ket{\Phi^-}$ will converge to the target state.
The control field varies smoothly and steers the system gently to
the target state as shown in Fig.~\ref{fig2}.  Moreover convergence
is exponential and
\begin{equation}
\begin{split}
  |f(t)| &= \kappa |\Tr(i[\rho(t),\rho_d(t)] H_1)| \\
         &= \kappa \norm{i[\rho(t),\rho_d(t)] H_1} \\
         &\le \kappa \norm{i[\rho(t),\rho_d(t)]} \cdot \norm{H_1}
\end{split}
\end{equation}
shows that $f(t)$ is bounded and we can choose $\kappa$ to ensure that
$|f(t)|$ is sufficiently small and the approximations inherent in the
model remain valid.

The method can also be utilized to increase the entanglement in the
initial state, i.e, to prepare a maximally entangled state starting
with a partially entangled one.  More specifically, if the system
initial starts in the state $\ket{\psi_0}
=\lambda_1\ket{++}+\lambda_2\ket{--}$ then the control design
produces a control field that steers the system from this state to
the desired maximally entangled state $\ket{\Phi^+}$. Choosing
$\rho_d=\ket{\Phi^-}\bra{\Phi^-}$ instead, we can similarly prepare
$\ket{\Phi^-}$, and it can be verified that steering the state to
$\ket{\Psi^-}$ simply requires inverting the sign of the control
field. Thus, not only can we prepare a maximally entangled state,
but we can select which state we prepare.

If the coupling constants of the local Hamiltonian for the two atoms
are not exactly identical, e.g., if $H_{\rm local}=\eta J(X \otimes
I+kI\otimes X)$ then changing to the $X$-basis gives $H_1=\eta J
\diag(1+k,1-k,-1+k,-1-k)$, which transforms to
\begin{equation}
  \tilde{H}_1 = \eta J
  \begin{pmatrix}
   0 & 0 & 0 & 1-k\\
   0 & 0 & 1+k & 0\\
   0 & 1+k & 0 & 0\\
   1-k & 0 & 0 & 0
\end{pmatrix}.
\end{equation}
Thus for $k\neq 1$ we can also steer the system from the product states
$\ket{+-}$ or $\ket{-+}$ to the Bell state $\ket{\Phi^\pm}$, i.e., for
this two-atom model Lyapunov control can be used to prepare any of the
four Bell states.

One limitation of the scheme is that the initial state must be in
the subspace $\S$, for example, $\S=\Span\{\ket{++},\ket{--}\}$, for
the control to be effective. This is not a shortcoming of the
proposed control scheme, however, because we can see from the
structure of $\tilde{H}_0$ and $\tilde{H}_1$ that the control system
is decomposable, hence not controllable on the whole
space~\cite{D'Alessandro}. More specifically, the dynamics on the
orthogonal subspaces $\S$ and $\S^\perp$ are independent, and
subspace populations are conserved quantities. Thus, for the above
Hamiltonian, \emph{no} control exists that steers population from
subspace $\S$ to $\S^\perp$ and vice versa.

\subsection{Interaction control}

\begin{figure}
\includegraphics[width=\columnwidth]{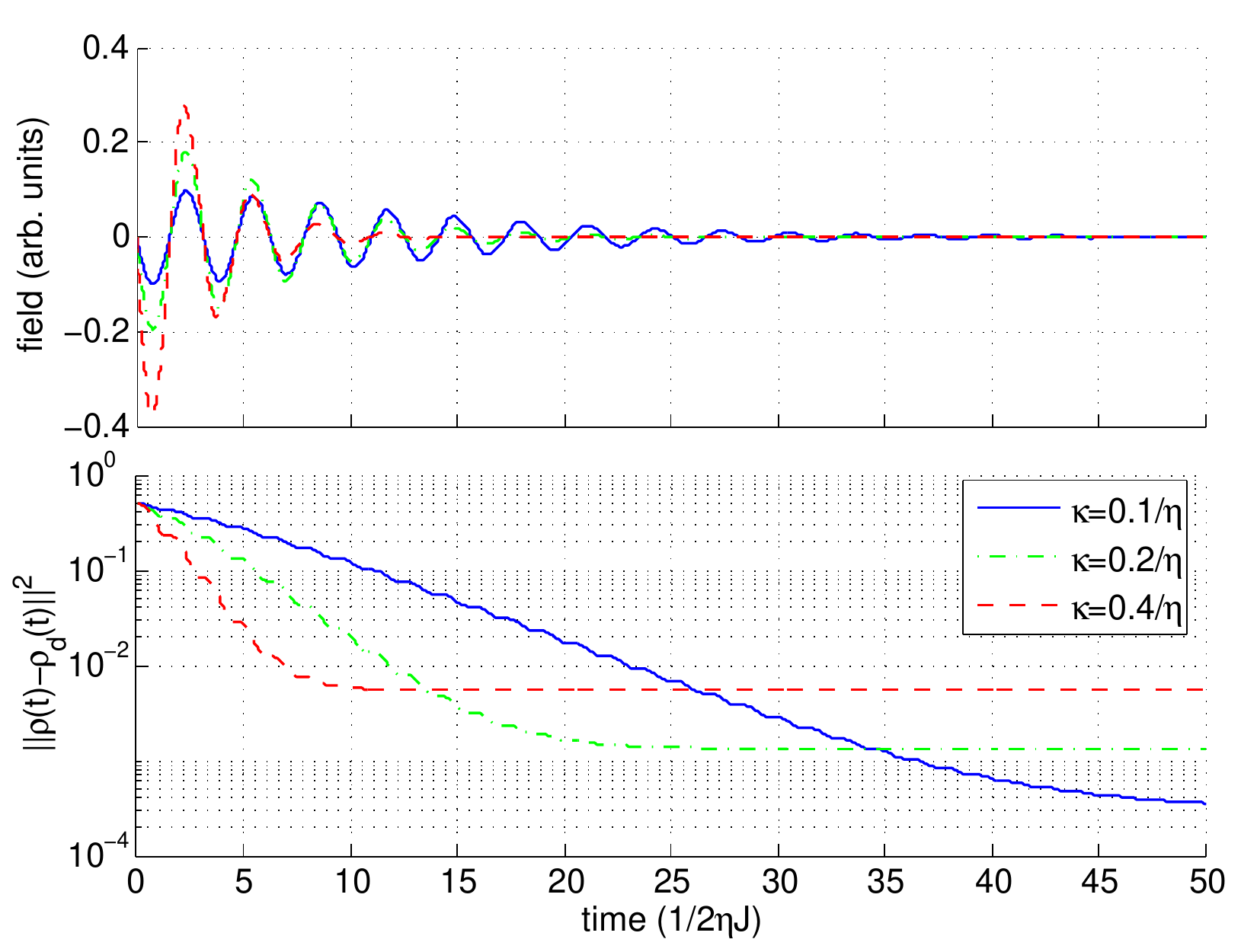}
\caption{Interaction control: control fields obtained from Lyapunov
design for different values of $\kappa$ and distance between the
system state and the target state with
$\rho_d(0)=\ket{\Psi^+}\bra{\Psi^+}$. The control design is robust
in that the field amplitude gently decreases to zero, and the
semilog distance plot shows that the convergence to the target state
is not only monotonic but actually exponential, although unlike in
the local control case, $||\rho(t)-\rho_d(t)||$ does not converge to
$0$. The final $\rho(t)$ is still maximally entangled with unit
concurrence.} \label{fig3}
\end{figure}

Instead of controlling the atoms locally, we can alternatively
control the nonlocal Hamiltonian $H_{\rm eff}$, if the underlying
physical system allows. In this case we choose $H_0=\eta J(X\otimes
I+ I\otimes X)$ and $H_1=2J(Z\otimes Z)$, or in the $X$-eigenbasis
\begin{align*}
H_0= 2\eta J\begin{pmatrix}
1 & 0 & 0 &  0 \\
0 & 0 & 0 &  0\\
0 & 0 & 0 &  0\\
0 & 0 & 0 &  -1
\end{pmatrix}, \quad
H_1=2J\begin{pmatrix}
0 & 0 & 0 &  1 \\
0 & 0 & 1 &  0\\
0 & 1 & 0 &  0\\
1 & 0 & 0 &  0
\end{pmatrix}.
\end{align*}
The Bell states are no longer the eigenstates of $H_0$. Hence, for
$\rho_d(0)=\ket{\Phi^+}\bra{\Phi^+}$, the target state is also
evolving with time, but for $\rho(0)=\ket{++}$ the dynamics is still
confined to the subspace $\S$ spanned by $\ket{++}$ and $\ket{--}$.
Therefore, the dynamics can again be reduced to a 2D subspace on
which we have
\begin{align*}
H_0= 2\eta J\begin{pmatrix}
1 & 0 \\
0 & -1
\end{pmatrix}, \quad
H_1=2J\begin{pmatrix}
0 & 1 \\
1 & 0
\end{pmatrix},
\end{align*}
as well as
\begin{align*}
\rho(0)=\begin{pmatrix}
1 & 0 \\
0 & 0
\end{pmatrix}, \quad
\rho_d(0)=\frac{1}{2}\begin{pmatrix}
1 & 1 \\
1 & 1
\end{pmatrix},
\end{align*}
where the orbit of $\rho_d(t)$ is the equator of the Bloch sphere.

\begin{figure}
\includegraphics[width=\columnwidth]{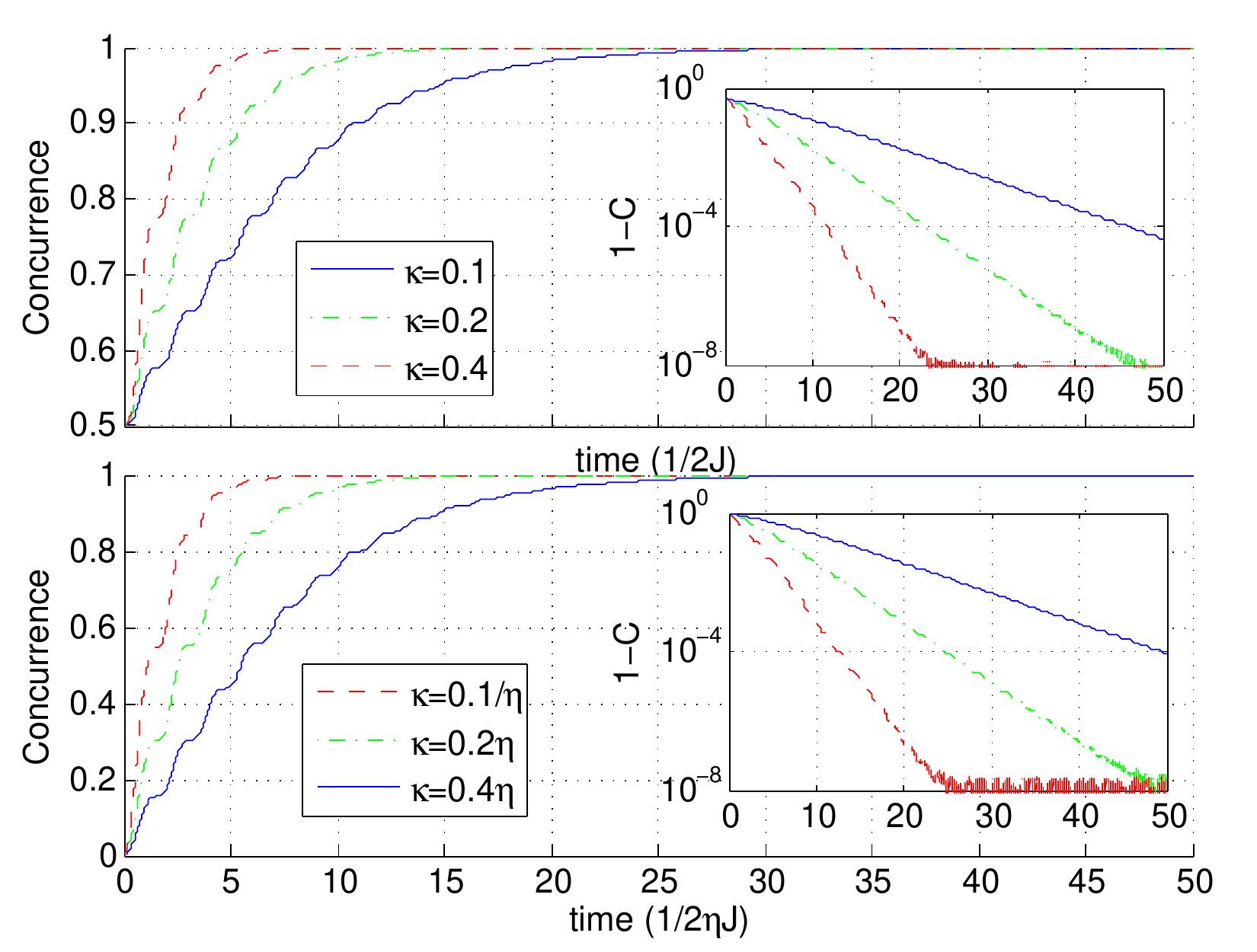}
\caption{Evolution of concurrence $\cal C$ under Lyapunov control for
different values of $\kappa$ for local control (top) and interaction
control (bottom) shows monotonic convergence to $1$.  Insets show the
error, i.e., $1-\C$ decreases effectively exponentially.}  \label{fig4}
\end{figure}

From the analysis in \cite{Wang-Schirmer} we can conclude that all
solutions in $\S$ will converge to the equator of the Bloch sphere,
i.e., states of the form
\begin{align*}
\rho=\frac{1}{2}\begin{pmatrix}
1 & e^{-i\alpha} \\
e^{i\alpha} & 1,
\end{pmatrix}
\end{align*}
which corresponds to the LaSalle invariant set $E$ of the original
problem satisfying
\begin{align*}
\rho=\frac{1}{2}\begin{pmatrix}
1 & 0 & 0 &  e^{-i\alpha} \\
0 & 0 & 0 &  0\\
0 & 0 & 0 &  0\\
e^{i\alpha} & 0 & 0 &  1.
\end{pmatrix}
\end{align*}
Thus, we can no longer guarantee $\rho(t)\rightarrow \rho_d(t)$ as
$t\to +\infty$, i.e., that the state converges to a particular Bell
state. This is illustrated in Fig.~\ref{fig3}, which shows that the
distance from the target state still decreases monotonically and
exponentially but the asymptotic value of $V(\rho(t),\rho_d(t))$ for
$t\to\infty$ now depends on $\kappa$ and is generally larger than
zero. However, since all the states in the set to which $\rho(t)$
converge are maximally entangled, we can still steer the system to a
maximally entangled state, and the concurrence still increases
monotonically to one (see Fig.~\ref{fig4}) but the relative phase
$\alpha$ of the state we converge to now depends on the exact
initial state and the feedback strength $\kappa$.

Strictly speaking, as the control $f(t)$ reduces to zero, the norm
of $H_{\rm local}$ will cease to be significantly smaller than that
of $H_{\rm eff}$, rendering the approximations made in the
derivation of $H_{\rm eff}$ invalid, unless we reduce the strength
of $H_{\rm local}$ accordingly.  However, in practice, the system
should already have reached a state with significant entanglement
before the model becomes invalid.\\

\section{Conclusion}

We have shown how to apply Lyapunov control to the problem of
generating entanglement between two distant two-levels atoms in
cavities connected by optical fibers. Given the Lyapunov control
design, despite the fact that the sufficient condition for a target
state to be asymptotically stable is not satisfied on the whole
state space, we can still ensure it is almost globally
asymptotically stable on certain subspace. Therefore, within that
subspace we can drive the system from a product state to a maximally
entangled state. The Lyapunov control design has the advantage of
much greater robustness compared to simple geometric schemes, and
optimality in the sense that the distance from the maximally
entangled target state is monotonically decreasing, and the
convergence speed is exponential. We have discussed two control
paradigms: to control the local Hamiltonian, as well as to control
the effective interaction Hamiltonian between the two atoms. In both
cases we can generate a maximally entangled state from an initial
product state: for the formal case the system state will converge to
a stationary Bell state, while for the latter case the relative
phase of the final state will keep varying under the Hamiltonian,
since the target state is non-stationary. Moreover, in the latter
case, the model becomes invalid when the control amplitude becomes
sufficiently small. Therefore, the former control paradigm is
preferable.  The Lyapunov control design can be also used to steer
partially entangled states to a maximally entangled state, however,
the control is only effective for initial states in the subspace
where the target state is asymptotically stable. This is not a
limitation of the control design, however, but a consequence of the
fact that the controlled system is decomposable into two orthogonal
subspaces on each of which the dynamics is invariant. In this sense,
the Lyapunov control design is as effective as is possible within
the constraints of the model.

\acknowledgments

We gratefully thank the stimulating discussions with Prof. Sougato
Bose. SGS acknowledges funding from EPSRC ARF Grant EP/D07195X/1,
Hitachi, and NSF Grant PHY05-51164.

\end{document}